\documentclass[preprintnumbers]{revtex4}
\usepackage{eurosym}
\usepackage{amsmath}
\usepackage{amssymb}
\usepackage{graphicx}
\usepackage{color}
\usepackage{epstopdf}
\usepackage{epstopdf}
\usepackage{bm}

\DeclareUnicodeCharacter{2212}{-}
\begin{document}

\date{\today }
\title{Intervalley Magnetotrions Tunable by Electric and Magnetic Fields in Buckled Two-Dimensional Materials 
}
\author{Roman Ya. Kezerashvili$^{1,2}$, Shalva M. Tsiklauri,$^{3}$ and
Anastasia Spiridonova$^{1}$}
\affiliation{\mbox{$^{1}$New York City College of Technology, The City University of New
York, USA} \\
$^{2}$The Graduate School and University Center, The City University of New
York, USA\\
$^{3}$Borough of Manhattan Community College, The City University of New
York, USA}

\begin{abstract}

We develop a theoretical framework for intervalley magnetotrions in buckled two-dimensional materials, including silicene, germanene, and stanene, subjected to perpendicular electric and magnetic fields. Within the effective-mass approximation, the three-particle Schr\"odinger equation is formulated with the Rytova--Keldysh interaction potential and analyzed in the high-magnetic-field regime. We demonstrate that intervalley trions with equal electron and hole effective masses constitute an exceptional case for which the center-of-mass and internal motions separate exactly. The center-of-mass motion is governed by a two-dimensional harmonic-oscillator Hamiltonian, leading to quantized Landau states whose energies form electrically tunable Landau surfaces controlled by the magnetic field and the electric-field dependence of the carrier effective masses. The internal motion is investigated by solving the three-body Schr\"odinger equation within the framework of the hyperspherical harmonics method. Numerical calculations reveal that the trion binding energy increases monotonically with both magnetic and electric fields owing to the combined effects of magnetic confinement and electric-field-induced enhancement of the effective masses. The strongest binding is obtained for silicene, followed by stanene and germanene. The present work provides a unified description of both the collective center-of-mass motion and the internal dynamics of magnetotrions in Xene monolayers, demonstrating that both degrees of freedom can be independently manipulated by external electric and magnetic fields.


\end{abstract}

\maketitle

\section{Introduction}

Trions—bound states composed of an exciton and an additional electron or hole, known as negatively or positively charged excitons (X$^{\mp}$)—was predicted by Lampert in 1958 \cite{Lampert58}. This pioneering idea sparked decades of theoretical and experimental investigations into trions in bulk semiconductors, quantum well systems, and, more recently, two-dimensional (2D) materials. Only in 1993 Kheng \textit{et al.} \cite{Kheng1993} reported the observation of negatively charged excitons $X^{−}$ in semiconductor quantum wells, although $X^{+}$ cannot be formed in bulk materials \cite{FilikhinKezPLA1918,FilikhinKez} and was not experimentally detected. 

In the new millennium, among 2D materials, atomically thin transition metal dichalcogenides (TMDCs) have attracted particular interest due to their remarkable optical and electronic properties \cite{Kormanyos, RMP}. Since the first experimental observation of trions in monolayer MoS$_2$ in 2013 \cite{MoS23Heinz}, there has been a surge of research focused on understanding trionic states in TMDC monolayers and numerous experimental groups reported $X^{\mp}$ trions binding energies in the range of 20 to 40 meV (see reviews \cite{RMP,Durnev2018,Kezerashvili2019,Suris2022,Kezerashvili2026} and references herein).

Theoretical studies of trions have expanded rapidly following the discovery of monolayer TMDCs with large trion binding energies~\cite{Kezerashvili2026}. A broad spectrum of approaches has been employed, including variational methods \cite{Berkelbach2013,SeminaVar2017,Variational2021}, 
Tamm--Dancoff approximation \cite{ZhumagulovTamm,ZhumagulovTamm2}, stochastic variational calculations \cite{VargaNano2015,VargaPRB2016,Varga2020}, diffusion and path-integral Monte Carlo simulations \cite{DenFuncTheoryPIMC,Saxena,BerkelbachDifMonteCarlo,Drummond2016Diffusion,Szyniszewski,Szyniszewski2}, \textit{ab initio} many-body calculations \cite{Deilmann2017}, time-dependent density-matrix functional theory \cite{TimeDepdensity matrix functional theory}, finite-element methods \cite{model}, exact diagonalization \cite{Fey2020}, efficient treatments of the Coulomb interaction \cite{KumarPRB2025}, and few-body methods based on the Faddeev equations and hyperspherical harmonics \cite{Frederico2023,FilikhinKez,KezFew2017,KezPRB109trion}. These approaches have produced highly accurate trion binding energies in excellent agreement with experiment.

The theory of Mott excitons in strong magnetic fields was developed and extensively studied in the 1960s \cite{Elliott,Hasegawa1961,Shinada1965,Akimoto,Gorkov1965}. A key aspect of the theoretical description is the separation of the center-of-mass ($c.m.$) and internal motions of the electron--hole pair, as established in Refs.~\cite{Elliott,Gorkov1965,Lozovik,Shinada1965,Akimoto,Herold,Avron1978MagField,Dzyubenko1993}. For excitonic systems with unequal electron and hole effective masses, however, such a separation is generally impossible, and the $c.m.$ and internal motions become coupled in the presence of a magnetic field, even if the electron--hole interaction is translation invariant  \cite{Gorkov1965,Avron1978MagField,Dzyubenko2000PRL}.

Magnetoexcitons in TMDCs, Xenes, phosphorene, transition metals trichalcogenide monolayers, bilayers, and van der Waals heterostructures have been extensively investigated \cite{Ludwig,Tsaran2014,Macneill,Striv,Plechinger,Stier_2016,Zarena2018,Arora2018,Efimkin2018,Koperski,Gor2019,Xuan2020,Stier2018,Liu2019,Spiridonova2020,Muoi2020,KezerashviliSpir2021,KezSpir2022Xenes,KezSpirphosp2022,Kezerashvili2022TMTC} and references therein. Most theoretical studies focus on the internal motion of the electron--hole pair.

Theoretical studies of trions in semiconductor quantum wells, quantum dots, and two-dimensional materials subjected to magnetic fields have also attracted considerable attention, with numerous calculations devoted to the magnetic-field dependence of trion energy levels \cite{Stebe1987,Hawrylak1995,MacDonald1996,Whittaker1997,Chapman1998,Hawrylak1998,Szlufarska1999,Hawrylak1999,Stebe2000,Rivavarga12001,Rivavarga22001,Dzyubenko2000PRL,Dzyubenko2001,Kossut2002,Wojs2007PRBTrionED,Zarena2018,Chang2022Variational,Shelych,Aleksandrov2024}. We have cited these works, but the body of relevant literature is not limited to these publications. As early as the mid-1990s, it was predicted that $X^{-}$ trions are long-lived fermionic quasiparticles whose energy spectra exhibit Landau-level structures \cite{Hawrylak1995,MacDonald1996}.

Nowadays, theoretical studies of trions in magnetic fields are primarily focused on TMDC monolayers \cite{Chang2022Variational,Shelych,Aleksandrov2024}. It is important to emphasize that a magnetic field couples the c.m.\ and internal motions of the three-particle system, even in the absence of interparticle interactions. In contrast to TMDCs, no theoretical studies of trions in Xene monolayers under magnetic fields have been reported. The present work aims to fill this gap. The principal novelty of this work is the identification of intervalley magnetotrions in buckled two-dimensional materials as a rare charged three-body system that allows for the exact separation of the center-of-mass and internal motions under the equal-mass condition in a uniform high magnetic field. This exact treatment leads to the prediction of electrically tunable $c.m.$ Landau states and provides a unified theoretical description of both the collective and internal dynamics of magnetotrions under external electric and magnetic fields.

In this work, we investigate the effects of electric and magnetic fields on trions in Xene monolayers by calculating the electric- and magnetic-field dependence of the trion binding energy. The trion is treated as a three-particle system described by the Schr\"odinger equation, with the magnetic field incorporated through the vector potential in the symmetric gauge, while the electric field is applied perpendicular to the Xene monolayer. Our formalism enables the separation of the $c.m.$ and internal motions of the three-particle system. The $c.m.$ motion is described by a harmonic-oscillator-type equation, yielding Landau surfaces as functions of the electric and magnetic fields. The internal wave function is obtained by solving the three-body Schr\"odinger equation within the framework of the hyperspherical harmonics method.

The paper is organized as follows. In Sec. \ref{TheoryXene}, we present the low-energy model and intravalley/inter\-valley trions in Xene monolayers. The Schr\"{o}dinger equation for Mott-Wannier trions in electric and magnetic fields in Xene monolayers is presented in Sec. \ref{TheoryTrion}. Sec. \ref{C.M.Trion} presents a discussion for center-of-mass separation of composite systems in homogeneous magnetic fields. 
Separation of the $c.m.$ for non-interacting three charged particle of equal masses in a magnetic trap are considered in Sec. \ref{Non_Interacting}. Here we present the quantized Landau surfaces for trions in Xene monolayer. In Sec.~\ref{InternalMotion}, we investigate the internal motion of trions in Xene monolayers within the framework of the hyperspherical harmonics method. Using the Schr\"odinger equation with the Rytova--Keldysh interaction potential~\cite{Ritova,Keldysh}, the three-body problem is reduced to a system of coupled differential equations for the hyperradial functions. Numerical solutions of this system provide the binding energy 
for internal motion of trions in Xene monolayers. Concluding remarks follow in Sec. \ref{Conclusion}.

\section{Low-energy model and intravalley/inter\-valley trions in Xene monolayers}

\label{TheoryXene}

\begin{figure}[h!]
\centering
\includegraphics[width=17.0cm]{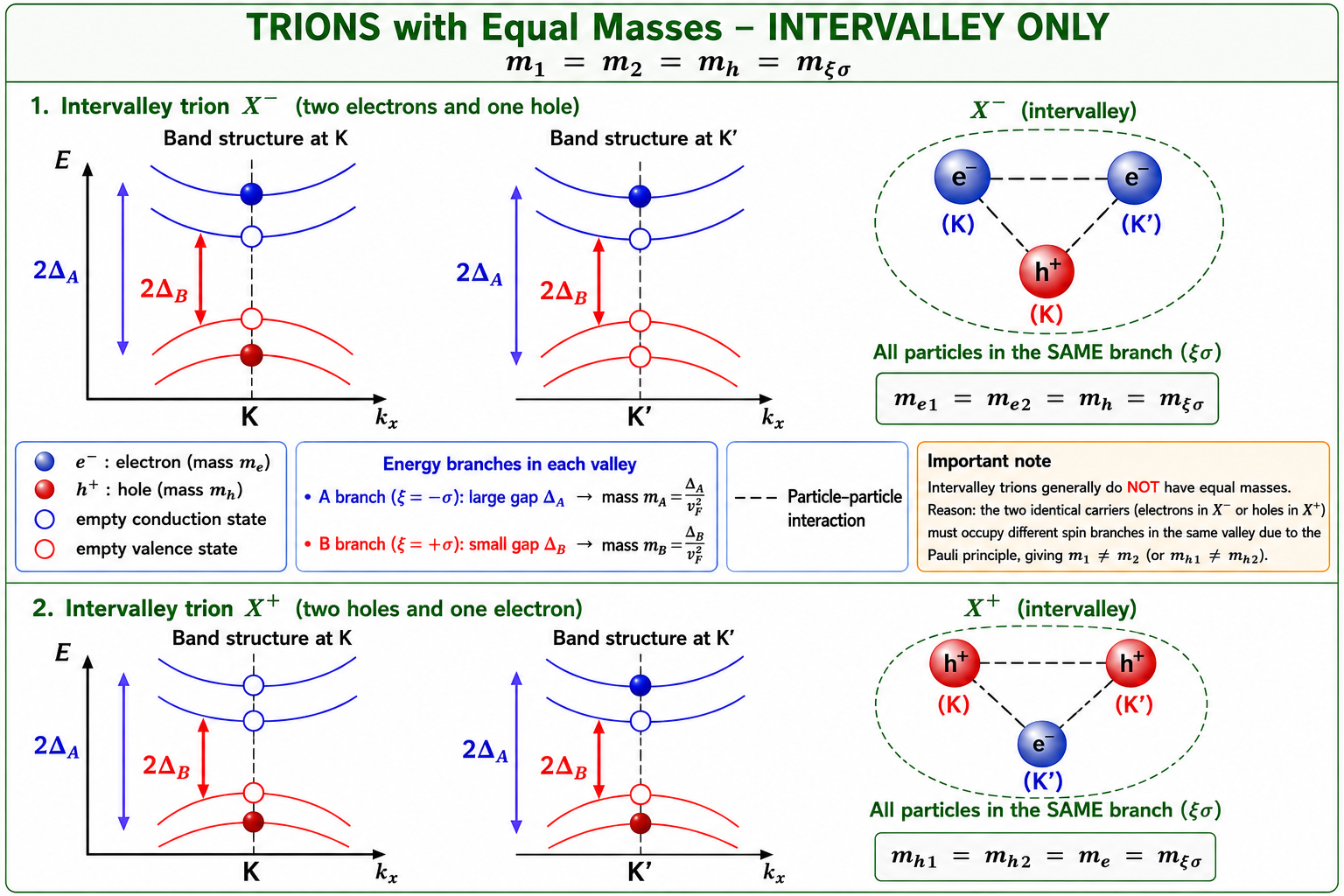}

\caption{(Color online) 
Schematic illustration of equal-mass intervalley trions in Xene monolayers. The upper and lower panels show negatively charged, \(X^{-}\), and positively charged, \(X^{+}\), trions, respectively. The band structures near the \(K\) and \(K^{\prime}\) valleys are shown together with the occupation of the conduction and valence bands. 
Although the constituent particles occupy different valleys (\(K\) and \(K^{\prime}\)), they belong to the same valley-spin branch \((\xi\sigma)\), yielding equal effective masses \(m_{e1}=m_{e2}=m_h=m_{\xi\sigma}\) for \(X^{-}\) and \(m_{h1}=m_{h2}=m_e=m_{\xi\sigma}\) for \(X^{+}\). The corresponding band occupations and three-body configurations are illustrated together with the \(A\)- and \(B\)-branch energy gaps. 
}
\label{Trion_schema}
\end{figure}

We start by outlining the low-energy model that describes excitonic complexes in Xene monolayers subjected to a perpendicular external electric field. Silicene, germanene, and stanene are buckled two-dimensional materials with a honeycomb lattice similar to graphene. Unlike graphene, however, the two triangular sublattices are vertically displaced with respect to the monolayer plane by a distance $d_0$, referred to as the buckling constant.

The buckled geometry makes Xenes highly sensitive to an electric field applied perpendicular to the monolayer. Such a field induces an on-site potential difference between the two sublattices and modifies both the band gap and the effective masses of the charge carriers. Near the $K/K^{\prime}$ points, the low-energy electronic structure is described by the Hamiltonian \cite{Tabert} ($\hslash = c = 1$)

\begin{equation}
\hat{H}
=
v_F
\left(
\xi p_x \hat{\tau}_x
+
p_y \hat{\tau}_y
\right)
-
\xi \Delta_{\rm gap}
\hat{\sigma}_z
\hat{\tau}_z
+
\Delta_z
\hat{\tau}_z ,
\label{eq:taberthamiltonian}
\end{equation}
where $v_F$ is the Fermi velocity, $\xi=\mp1$ labels the $K$ and $K^{\prime}$ valleys, and $\sigma=\mp1$ denotes the spin projection. The matrices $\hat{\tau}$ and $\hat{\sigma}$ are Pauli matrices acting in pseudospin and spin spaces, respectively. The intrinsic band gap is $2\Delta_{\text{gap}}$, and the electric-field-induced contribution is given by $\Delta_z = e d_0 E_{\perp}$, where $E_{\perp}$ is the electric field strength. The first term in Eq.~\eqref{eq:taberthamiltonian} corresponds to the graphene-like low-energy Hamiltonian \cite{CastroNeto2009a, Abergel2010}, the second term accounts for spin-orbit coupling, and the third term represents the sublattice potential difference induced by the electric field \cite{Drummond2012, Ezawa1, Ezawa2, Ezawa3}. 
The corresponding energy dispersion relation is $E(k)
=
\sqrt{
\Delta_{\xi\sigma}^{\,2}
+
v_F^2 p^2
},$
with the field-dependent half-gap

\begin{equation}
\Delta_{\xi\sigma}
=
\left|
\xi \sigma \Delta_{\rm gap}
-
e d_0 E_{\perp}
\right|.
\label{eq:deltaez}
\end{equation}

The application of a perpendicular electric field splits the conduction and valence bands into two branches. The larger gap corresponds to the $A$ exciton branch, while the smaller gap corresponds to the $B$ exciton branch. At the critical field
\begin{equation}
E_c
=
\frac{\Delta_{\rm gap}}
     {e d_0},
\end{equation}
one of the valley-spin branches becomes gapless, producing a Dirac-like spectrum at the $K/K^{\prime}$ points \cite{BBKPRB2019}.
Near the band extrema, the dispersion relation is approximately parabolic, yielding the effective mass

\begin{equation}
m_{\xi\sigma}
=
\frac{
\left|
\xi \sigma \Delta_{\rm gap}
-
e d_0 E_{\perp}
\right|
}
{v_F^2}.
\label{eq:effmassEz}
\end{equation}

When an electron is excited across the band gap, a hole is left in the valence band. The Coulomb attraction between the electron and hole can produce a neutral exciton. In doped systems, an exciton may bind an additional electron or hole, forming a negatively or positively charged trion, $X^{-}$ or $X^{+}$. Trions are therefore three-body bound states consisting of two identical charge carriers and one carrier of opposite charge.

In Xenes, excitons and trions belong to the Wannier--Mott class because their characteristic size is much larger than the lattice constant. Consequently, they can be described within the effective-mass approximation, which treats quasiparticles as particles with field-dependent effective masses interacting through screened Coulomb potentials. This approach has been widely employed in studies of excitonic complexes in two-dimensional materials
\cite{Kormanyos,BekReichman,Kezerashvili2019,Suris2022,Kezerashvili2026,RMP}. 
The valley-spin structure of Xenes leads to several possible trion configurations. Trions may be classified as intravalley or intervalley complexes depending on the valleys occupied by their constituent particles.

For an intravalley trion, all particles occupy the same valley, $K$ or $K^{\prime}$, but the two identical carriers must satisfy the Pauli principle. Depending on the spin configuration and the valley-spin structure of the Xene, the identical particles may have to occupy different spin branches, which can correspond to different effective masses. Intervalley trions involve carriers occupying different valleys. Depending on the valley-spin configuration, they may possess either equal or unequal masses. Equal masses arise when all particles belong to branches characterized by the same value of $\xi\sigma$, so that $\Delta_{\xi_1\sigma_1}
=
\Delta_{\xi_2\sigma_2}
=
\Delta_{\xi_3\sigma_3}$
which leads to

\begin{equation}
m_1=m_2=m_3.
\end{equation}
This mass degeneracy represents the central symmetry condition of the present work, providing the mathematical foundation that enables the exact decoupling of the collective and internal degrees of freedom of the intervalley magnetotrons in a uniform high magnetic field.

In contrast, unequal-mass intravalley and intervalley trions occur when carriers occupy different valley-spin branches, such as the $A$ and $B$ branches. The corresponding effective masses $m_A \neq m_B$. 
Such systems constitute genuine unequal-mass three-body problems. Their binding energies, spatial structure, and magnetic-field response depend sensitively on the electric field through the field-dependent effective masses given by Eq.~(\ref{eq:effmassEz}). Equal-mass intervalley trions are particularly attractive from a theoretical perspective because the corresponding three-body Hamiltonian possesses a higher degree of symmetry.

\section{Theoretical model for trions in electric and magnetic fields }

\label{TheoryTrion} 

The behavior of trions in two-dimensional materials is strongly modified by the application of external magnetic and electric fields. In the presence of a perpendicular magnetic field, the electron and hole experience additional magnetic confinement, leading to the formation of magnetotrions. While for Xene monolayers, the perpendicular electric field leads to the manipulation of the gap, and, therefore, the effective masses of electrons and holes. 
In order to obtain the eigenfunctions and eigenenergies of a 2D trion in Xenes, when the electric and magnetic fields are perpendicular to the Xene monolayer, we write the Schr\"{o}dinger equation for an interacting
three-particle electron-hole system. Because we are considering the varying
electric field $E_{\perp }$, which is directed along $z$-axis, the
corresponding term in the 2D Schr\"{o}dinger equation vanishes. However, the
effect of the electric field action is present through the effective mass as
follows from Eq. (\ref{eq:effmassEz}). Thus, one can write 2D Schr\"{o}%
dinger equation for the interacting three-particle system in an external
magnetic field within the effective mass approximation in the following form 
\cite{Zarena2018,Chang2022Variational,Shelych}:
\begin{equation}
\left[ \overset{3}{\underset{i=1}{\sum }}\frac{1}{2m_{i}}\left( \mathbf{p}%
_{i}-{q_{i}}\mathbf{A}_{i}\right) ^{2}+\overset{3}{\underset{i<j}{%
\sum }}V_{ij}(\left\vert \mathbf{r}_{i}-\mathbf{r}_{j}\right\vert )\right]
\Psi (\mathbf{r}_{1},\mathbf{r}_{2},\mathbf{r}_{3})=\mathcal{E}\Psi (\mathbf{r}_{1},%
\mathbf{r}_{2},\mathbf{r}_{3}),  \label{MagTrion}
\end{equation}%
with $\mathbf{p}_{i}=-i\mathbf{\nabla }_{i}$, $q_{i}$ and $m_{i}$
the charge and effective mass of particle $i$, respectively, and $\mathbf{A}_{i}=\mathbf{r}%
_{i}\times \frac{\mathbf{B}}{2}$ is the vector potential at the position of
the particle in the symmetric gauge corresponding to the uniform
perpendicular external magnetic field $\mathbf{B}=(0,0,B)$, and $%
V_{ij}(\left\vert \mathbf{r}_{i}-\mathbf{r}_{j}\right\vert )$\ is a 2D
screened electrostatic interaction. 
For a negative trion $X^{-}$, $
q_1=q_2=-e$ and $q_3=+e$,   
where particles 1 and 2 are electrons and particle 3 is a hole. Conversely, for a positive trion $X^{+}$, $
q_1=q_2=+e$ and $q_3=-e$. 

In Eq. (\ref{MagTrion}), several terms are omitted. Interband coupling is neglected because the excitations are assumed to occur near the band extrema, where the conduction and valence bands can be approximated by parabolic dispersions. 
Zeeman contributions are likewise omitted because they do not affect the binding energies or structural properties of excitonic complexes \cite{Zarena2018}. Valley-Zeeman terms are also neglected unless the valley splitting is comparable to the trion binding energy. 
The finite thickness of the monolayer and lattice-scale corrections are neglected because trions in Xenes are treated as Wannier--Mott complexes whose spatial extent is larger than the lattice constant. Finally, Eq.~(\ref{MagTrion}) is written in the laboratory-particle coordinates; consequently, the coupling between the center-of-mass and relative motions is not shown explicitly. These coupling terms arise after transformation to Jacobi coordinates and must generally be retained, except under special conditions, such as equal effective masses together with the appropriate pseudomomentum treatment, where a partial or complete separation of the motions becomes possible.

The 2D screened electrostatic
interaction is described by the Rytova--Keldysh potential \cite%
{Ritova,Keldysh}.\ The Rytova-Keldysh potential describes the Coulomb
interaction screened by the polarization of the electron orbitals in the 2D
lattice and has the following form

\begin{equation}
V_{ij}(\left\vert \mathbf{r}_{i}-\mathbf{r}_{j}\right\vert )=\frac{\pi
kq_{i}q_{j}}{2\kappa \rho _{0}}\left[ H_{0}\left( \frac{\left\vert \mathbf{r}%
_{i}-\mathbf{r}_{j}\right\vert }{\rho _{0}}\right) -Y_{0}\left( \frac{%
\left\vert \mathbf{r}_{i}-\mathbf{r}_{j}\right\vert }{\rho _{0}}\right) %
\right] ,  \label{Keldysh}
\end{equation}%
where $r=\left\vert \mathbf{r}_{i}-\mathbf{r}_{j}\right\vert $ is the
relative coordinate between two charge carriers $q_{i}$ and $q_{j}$. In Eq.~(%
\ref{Keldysh}) $k=9\times 10^{9}$ N$\cdot $m$^{2}$/C$^{2}$, $\kappa $ is the
dielectric constant of the environment that is defined as $\kappa
=(\varepsilon _{1}+\varepsilon _{2})/2$, where $\varepsilon _{1}$ and $%
\varepsilon _{2}$ are the dielectric constants of two materials that the
Xenes layer is surrounded by either below or/and above the monolayer, $\rho _{0\text{ }}$ is the screening length, which sets the boundary
between two different behaviors of the potential due to a nonlocal
macroscopic screening, and $H_{0}(\frac{r}{\rho _{0}})$ and $Y_{0}(\frac{r}{%
\rho _{0}})$ are the Struve function and Bessel function of the second kind,
respectively. The screening length $\rho _{0}$ can be written as $\rho
_{0}=(2\pi \chi _{2D})/(\kappa )$ \cite{Berkelbach2013}, where $\chi _{2D}$
is the 2D polarizability, which in turn is given by $\chi _{2D}=l\varepsilon
/4\pi $~\cite{Keldysh}, where $\varepsilon $ is the bulk dielectric constant
of the Xene monolayer. For large distances $\ r>>\rho _{0\text{ }}$the
potential has the three-dimensional bare Coulomb tail $V_{ij}(r\mathbf{)=}%
\frac{kq_{i}q_{j}}{\epsilon r}$, while at very small distances, smaller than
the screening length $r<<\rho _{0\text{ }}$, it becomes a logarithmic
potential like a potential of a point charge in two dimensions: $V_{ij}(r%
\mathbf{)=}\frac{kq_{i}q_{j}}{\epsilon \rho _{0\text{ }}}\left[ \ln \left( 
\frac{r}{2\rho _{0}}\right) +\gamma \right] $, where $\gamma $ is the Euler
constant. Therefore, the potential (\ref{Keldysh}) becomes the standard bare
Coulomb potential at $r>>\rho _{0\text{ }}$ and diverges logarithmically at $%
r<<\rho _{0\text{ }}.$ A crossover between these two regimes takes place
around distance $\rho _{0}$. Thus, at small distances between charge
carriers the short-range interaction strength decreases, while the
long-range interaction strength is unaffected and is the bare
three-dimensional Coulomb potential. It is worth noting that in Ref. \cite%
{Rubio} a very good approximation to the RK potential that is simpler to
use, fairly precise in both limits and remarkably accurate for all distances
was introduced.

Let us consider trions in high magnetic field. The distinction between weak- and high-magnetic-field regimes for trions is determined by the competition between magnetic confinement and the Coulomb interaction. A convenient criterion is obtained by comparing the magnetic length,

\begin{equation}
l_B=\sqrt{\frac{1}{eB}},
\end{equation}
with the characteristic trion radius \(a_T\). The high-field regime is reached when the magnetic length becomes comparable to or smaller than the trion size, $l_B \lesssim a_T,$
so that magnetic confinement significantly modifies the spatial structure of the trion. An equivalent criterion is obtained by comparing the cyclotron energy, 
$\omega_c={ eB}/{m^*}$, where $m^*$ is an effective mass, 
%
with the trion binding energy \(E_b\). The magnetic field is considered strong when
%
$\omega_c \gtrsim E_b$,
%
indicating that Landau quantization plays an important role in the trion dynamics. For Xene monolayers, where trion binding energies are typically of the order of several tens of meV and the characteristic trion radius is a few nanometers, these conditions are generally satisfied at magnetic fields of approximately \(15\!-\!20\) T. Therefore, magnetic fields \(B \gtrsim 15\) T are commonly regarded as the onset of the high-field regime for trions, while fields above \(20\!-\!30\) T correspond to a regime in which magnetic confinement strongly influences the trion spectrum and wave functions. 

The magnetic-field range considered in the present work is experimentally accessible using modern pulsed high-field facilities at the National High Magnetic Field Laboratory in Los Alamos, New Mexico. High-field magneto-optical experiments on monolayer transition-metal dichalcogenides have investigated excitonic properties in magnetic fields up to 65 T using magneto-reflection and magneto-absorption spectroscopy, enabling measurements of valley Zeeman splitting, exciton diamagnetic shifts, exciton radii, dielectric screening, reduced masses, binding energies, and excited Rydberg states \cite{Stier_2016,Stier2016_JVB,Stier2016_NanoLett,Stier2018_PRL}. More recently, ultrahigh magnetic fields reaching 91 T have enabled even more precise determination of exciton reduced masses, dielectric properties, binding energies, and Rydberg exciton spectra \cite{Goryca2019}. These developments demonstrate that the magnetic-field interval investigated in the present work (15–80 T) lies well within current experimental capabilities.

One can write the Hamiltonian for Eq. (\ref{MagTrion}) for the high magnetic field regime.  
For the ground \(s\)-state trion, \(\langle L_z\rangle=0\), and therefore the orbital term $\sum_{i=1}^{3}
\frac{q_iB}{2m}
l_{zi}$, where $l_z$ is the orbital momentum of the particle,  does not contribute to the energy. In the high-magnetic-field regime, the dominant magnetic effect arises from the diamagnetic confinement term proportional to \(B^2\), yielding the simplified Hamiltonian

\begin{equation}
H=
\sum_{i=1}^{3}
\left[
\frac{\mathbf{p}%
_{i}^2}{2m_i}
+
\frac{q_i^2B^2}{8m_i}r_i^2
\right]
+
\sum_{i<j}^{3}
V_{ij}(r_{ij}),
\label{eq:HighBTrion}
\end{equation}
where the second term describes the magnetic confinement of the charge carriers.

\section{Center-of-mass separation of composite systems in homogeneous magnetic fields}

\label{C.M.Trion}
\subsection{Neutral two-particle systems}

The problem of describing composite systems in an external magnetic field has long been a subject of considerable theoretical interest. One of the simplest and most fundamental cases is that of a neutral two-body system, for which the separation of the center-of-mass 
and internal motions are a prerequisite for obtaining an exact solution. In the presence of a magnetic field, however, this factorization is highly nontrivial because the vector potential couples the $c.m.$ and relative coordinates. Nevertheless, an exact separation was successfully achieved within the nonrelativistic framework through the introduction of the conserved magnetic pseudomomentum, establishing the theoretical foundation for the description of neutral composite systems in homogeneous magnetic fields.

In a pioneering work, Gor'kov and Dzyaloshinskii \cite{Gorkov1965} developed the quantum-mechanical theory of the Mott exciton in a homogeneous magnetic field by introducing the concept of the conserved magnetic pseudomomentum for a neutral electron--hole pair  
\begin{equation}
\hat{\mathbf{K}}
=
\mathbf{p}_e
+
\mathbf{p}_h
-
\frac{e}{2}\,
\mathbf{B}\times
\left(
\mathbf{r}_e-\mathbf{r}_h
\right),
\label{eq:K_lab}
\end{equation}
where $\mathbf{p}_e$
and 
$\mathbf{p}_h$
are the canonical momenta of the electron and hole, respectively.
They demonstrated that, despite the presence of the magnetic field, the center-of-mass and relative motions can be separated through an appropriate gauge transformation, leading to an effective Hamiltonian for the internal motion. Because for a neutral exciton in a homogeneous static magnetic field, the conservation of the pseudomomentum $\hat{\mathbf{K}}$ is expressed by the vanishing commutator with the Hamiltonian from Eq. (\ref{MagTrion}) written in the case of the exciton, $[H,\hat{\mathbf{K}}]=0$, the Hamiltonian and the pseudomomentum operator can be simultaneously diagonalized. Consequently, the eigenstates of the neutral exciton may be chosen as simultaneous eigenfunctions of both operators. While the center-of-mass coordinate $\mathbf{R}$ separates from the total wave function, the $c.m.$ motion represented by the conserved magnetic pseudomomentum $\mathbf{K}$, remains a parameter in the internal Hamiltonian, so that the exciton spectrum depends explicitly on the center-of-mass motion,
\begin{equation}
    E=E_n(\mathbf{K}).
\end{equation}
Different values of $\mathbf{K}$ therefore correspond to different internal energies and wave functions. Only for the special case
    $\mathbf{K}=0$,
which describes optically active excitons at rest, does the internal Hamiltonian become independent of the center-of-mass motion, resulting in a complete decoupling of the $c.m.$ and relative dynamics.

\subsection{Charged three-particle and general $N$-particle systems 
}

The problem of separating the center-of-mass and internal motions becomes considerably more complicated for charged few-body systems in a homogeneous magnetic field. A general theoretical framework was developed by Avron, Herbst, and Simon~\cite{Avron1978MagField}, who considered an arbitrary $N$-particle system interacting through translation-invariant potentials. They showed that, although the ordinary translational symmetry is broken by the coordinate-dependent vector potential, the Hamiltonian remains invariant under the group of \emph{magnetic translations}. The corresponding conserved quantity is the total magnetic pseudomomentum,
\begin{equation}
\hat{\mathbf{K}}
=
\sum_{i=1}^{N}
\left(\mathbf{p}_i
+
\frac{q_i}{2}\,
\mathbf{B}\times\mathbf{r}_i
\right),
\label{eq:KN}
\end{equation}
which satisfies
\begin{equation}
[H,\hat{\mathbf{K}}]=0.
\end{equation}
In the last equation, $H$ is the Hamiltonian of $N$ charged particles in a magnetic field.

The algebra of the pseudomomentum components is determined by the total charge,
\begin{equation}
Q=\sum_{i=1}^{N}q_i,
\end{equation}
and is given by \cite{Avron1978MagField}
\begin{equation}
[K_\alpha,K_\beta]
=
-iQ\,\epsilon_{\alpha\beta\gamma}B_\gamma.
\label{eq:KKgeneral}
\end{equation}

For neutral systems, $Q=0$, all components of $\mathbf{K}$ commute, $[K_\alpha,K_\beta]=0$, and may therefore be simultaneously diagonalized together with the Hamiltonian. Consequently, the Hilbert space decomposes into sectors labeled by the conserved pseudomomentum, allowing an exact separation of the $c.m.$ coordinate from the internal degrees of freedom. In contrast to the zero-field case, however, the reduced Hamiltonian generally depends parametrically on $\mathbf{K}$.

The situation is fundamentally different for charged systems, $Q\neq0$, including charged three-body complexes such as trions. In this case, the nonvanishing commutator in Eq.~(\ref{eq:KKgeneral}) prevents the simultaneous diagonalization of all components of the pseudomomentum. As a result, the Hamiltonian cannot be factorized into independent $c.m.$ and internal parts, and an exact separation of the $c.m.$ motion is impossible. Instead, only a partial reduction of the $c.m.$ degrees of freedom can be achieved by projecting the Hamiltonian onto Landau-level (harmonic-oscillator) subspaces. Consequently, the internal dynamics remains intrinsically coupled to the $c.m.$ motion through the magnetic field. This work established the general symmetry principles governing few-body systems in magnetic fields and provided the mathematical foundation for subsequent studies of trions and charged complexes in semiconductor nanostructures.

Following
Refs.~\cite{Dzyubenko2000SSC,Dzyubenko2001}, as an example, for the $X^{-}$ trion
in a uniform perpendicular magnetic field, the conserved magnetic-translation
(pseudomomentum) operator can be written as
%
$\hat{\mathbf K}
= \sum_{i=1}^{2}
\left[\mathbf{p}_{ie}
+ e\mathbf A(\mathbf r_{ie})
\right] + \mathbf{p}_{h}
- 
e\mathbf A(\mathbf r_h)$,
%
which satisfies
$[H,\hat{\mathbf K}]$ = 0.
However, for a charged trion with total charge
$Q=\sum_i q_i \neq 0$, the components of $\hat{\mathbf K}$ obey
the noncommutative algebra
\begin{equation}
[\hat K_x, \hat K_y]
=
- iQ B,
\end{equation} 
so that $\hat K_x$ and $\hat K_y$ do not commute, and so that $\hat K_x$ and $\hat K_y$ cannot be simultaneously
diagonalized.
Consequently, only $\hat{\mathbf K}^2$ is conserved and is used
to classify magnetotrion states. For $Q\neq 0$, this structure
leads to Landau-like quantization of the trion $c.m.$
motion and modifies the internal binding energy.

In Refs. 
\cite{Dzyubenko2000PRL,Dzyubenko2000SSC,Dzyubenko2001} were established a comprehensive theoretical framework for charged electron--hole complexes in strong magnetic fields. They demonstrated that, although the complete separation of the $c.m.$ and internal motions is impossible for charged systems because the components of the magnetic pseudomomentum do not commute, the exact symmetry associated with magnetic translations can nevertheless be preserved. By developing an operator formalism based on the conserved magnetic translation operator, Bogoliubov canonical transformations, and coherent-state techniques, they constructed a basis compatible with both translational and rotational symmetries, achieving a \emph{partial} separation of the $c.m.$ and internal degrees of freedom. The formalism introduces an additional exact quantum number, the oscillator quantum number $k$, which labels the macroscopically degenerate Landau states and remains conserved in magneto-optical transitions. It also yields exact optical selection rules for trions, explains the existence of optically inactive (``dark'') triplet trion states in translationally invariant systems, and provides an efficient and physically transparent framework for high-precision calculations of trions, and other charged Coulomb complexes. This series of works has become a cornerstone of modern theoretical descriptions of charged few-body systems in strong magnetic fields. 

More recently, the theory of two-dimensional magnetotrions has been revisited using modern formulations based on magnetic-translation symmetry and exact numerical treatments of charged three-body systems, providing a refined description of the Landau-level structure and magnetotrion spectra in strong magnetic fields \cite{Shelych,Aleksandrov2024}. These developments further emphasize the importance of the interplay between collective magnetic motion and internal Coulomb correlations in charged excitonic complexes.

Simonov \cite{Simonov2013} developed an exact formalism for separating the center-of-mass and internal motions of a neutral three-body system in a homogeneous magnetic field. He showed that exact factorization is possible when two particles are identical, $m_1=m_2=m$ and $e_1=e_2=-e$, while the third particle has the compensating charge $e_3=2e$  satisfying $e_1+e_2+e_3=0$, and an arbitrary mass $m_3$. Using Jacobi coordinates and a magnetic phase transformation, the Hamiltonian becomes fully separable, allowing exact analytical solutions for three charged particles in the absence of interactions between them. The formalism is applied to the helium atom and neutral baryons, providing a rigorous framework for studying neutral three-body systems in strong magnetic fields~\cite{Simonov2013,Simonov2019}. The important conclusion is that the center-of-mass coordinate can be removed,
but unlike the zero-field case, the reduced Hamiltonian depends explicitly on the
conserved pseudomomentum $\mathbf K$. Thus, the internal motion remains
parameterized by $\mathbf K$, producing magnetically modified energy bands.

Thus, the inability to separate the $c.m.$ and internal motions in a magnetic field originates from the breaking of ordinary translational symmetry by the coordinate-dependent vector potential. The above developments demonstrate that the possibility of separating the center-of-mass and internal motions is determined by the symmetry of the system in a magnetic field. For neutral systems, the conserved magnetic pseudomomentum enables an exact separation of the $c.m.$ coordinate, although the internal Hamiltonian, and consequently the energy spectrum, generally depends parametrically on the conserved pseudomomentum $\mathbf{K}$. In contrast, for charged systems only a partial separation is possible because the components of the magnetic pseudomomentum do not commute. These fundamental symmetry properties form the theoretical basis for the description of magnetoexcitons and trions in low-dimensional materials.

The quantum mechanical theories discussed above establish that for generic charged three-body systems, the $c.m.$ and internal motions cannot be separated exactly in a magnetic field. In contrast, the present intervalley magnetotrion in Xene monolayers in a high uniform magnetic field represents an exceptional charged three-body system that permits an exact separation of these degrees of freedom. This unique symmetry breaks the general restrictions of previous frameworks and forms the foundation of the theoretical model developed in the following sections.

\section{\protect\bigskip Non-interacting  three charged particles in a magnetic trap}

\label{Non_Interacting}
To establish the exact separation of the $c.m.$ and internal motions, we first analyze the simpler problem of three noninteracting particles with charges equal in magnitude, $|q_i| = e$ ($i = 1,2,3$), confined in a magnetic trap, where particles 1 and 2 carry charges of the same sign, while particle 3 carries the opposite sign, corresponding to a negatively charged ($X^-$) or positively charged ($X^+$) trion. The corresponding Hamiltonian in Cartesian  coordinates, with the angular momentum of
particle terms omitted that do not depend on coordinates, reads 
\begin{equation}
H_{0}=-\frac{1}{2m_{1}}\nabla _{1}^{2}-\frac{1}{2m_{2}}%
\nabla _{2}^{2}-\frac{1}{2m_{3}}\nabla _{3}^{2}+\frac{e^{2}B^{2}}{%
8m_{1}}r_{1}^{2}+\frac{e^{2}B^{2}}{8m_{2}}r_{2}^{2}+\frac{%
e^{2}B^{2}}{8m_{3}}r_{3}^{2}.
\label{H0}
\end{equation}

Each term in Hamiltonian $H_{0}$\ includes the mass of the particle. Let's find a way to separate the relative and the center-of-mass motion for three charged particles in magnetic field.

\begin{figure}[h!]
\centering
\includegraphics[width=12.0cm]{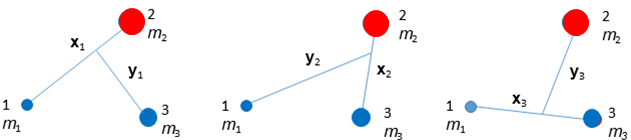}
\caption{(Color online) Schematics of partition trees of Jacobi coordinates when three charged particles have different masses. 
}
\label{Jacobi}
\end{figure}

As the first step, let us 
introduce sets of mass-scaled Jacobi
coordinates \cite{Avery,JibutiSh}. There are three equivalent sets of Jacobi
coordinates, and there is an orthogonal transformation between these sets 
\cite{FM,FilKez2024}. For three nonidentical particles that have different masses, the mass-scaled Jacobi coordinates 
for the partition $i$ read as follows \cite{Avery,JibutiSh,FM}: 
\begin{eqnarray}
\mathbf{x}_{i} &=&\sqrt{\frac{m_{j}m_{k}}{(m_{j}+m_{k})\mu }}(\mathbf{r}_{j}-%
\mathbf{r}_{k}),  \notag \\
\mathbf{y}_{i} &=&\sqrt{\frac{m_{i}\left( m_{j}+m_{k}\right) }{%
(m_{i}+m_{j}+m_{k})\mu }}\left( \frac{m_{j}\mathbf{r}_{j}+m_{k}\mathbf{r}_{k}%
}{m_{j}+m_{k}}-\mathbf{r}_{i}\right) ,\text{ \ }i\neq j\neq k=1,2,3,  \notag
\\
\mathbf{R} &=&\frac{m_{1}\mathbf{r}_{1}+m_{2}\mathbf{r}_{2}+m_{3}\mathbf{r}%
_{3}}{\sqrt{(m_{1}+m_{2}+m_{3})\mu }},  \label{Jacobi3}
\end{eqnarray}%
where 
\begin{equation}
\mu =\sqrt{\frac{m_{i}m_{j}m_{k}}{m_{i}+m_{j}+m_{k}}}\ \   \label{EffecMass}
\end{equation}%
is the three-particle effective mass. 
In Eqs. (\ref{Jacobi3}) the subscripts $i$, $j$, and $k$ are a cyclic
permutation of the particle numbers. Trees of Jacobi coordinates for a
three-particle system, when particles have different masses, are shown in
Fig. \ref{Jacobi}.

In Eq. (\ref{MagTrion}), the potential $V_{ij}(|\mathbf{r}_i-\mathbf{r}_j|)$ can be written using the Jacobi coordinates $\mathbf{x}_{k}$ for a particular partition $k$ 
as $V_{k}(|\mathbf{x}_k|)$ that depends on the corresponding coordinate set
which are expressed in terms of the chosen set of mass-scaled Jacobi
coordinates. The orthogonal transformation between three different sets of the Jacobi coordinates has the form \cite{FilKez2024,Filikhin2024PRD}:
\begin{equation}
\left(
\begin{array}{c}
\label{tran}\mathbf{x}_{i} \\
\mathbf{y}_{i}%
\end{array}%
\right) =\left(
\begin{array}{cc}
C_{ik} & S_{ik} \\
-S_{ik} & C_{ik}%
\end{array}%
\right) \left(
\begin{array}{c}
\mathbf{x}_{k} \\
\mathbf{y}_{k}%
\end{array}%
\right) ,\ \ C_{ik}^{2}+S_{ik}^{2}=1, \quad k\neq i,
\end{equation}%
where
\begin{equation*}
C_{ik}=-\sqrt{\frac{m_{i}m_{k}}{(M-m_{i})(M-m_{k})}},\quad S_{ik}=(-1)^{k-i}%
\mathrm{sign}(k-i)\sqrt{1-C_{ik}^{2}}.
\end{equation*}%
Here, $M$ is the total mass of the system. 

To separate the internal and collective degrees of freedom, we transform the Cartesian particle coordinates $\mathbf{r}_1$, $\mathbf{r}_2$, and $\mathbf{r}_3$ to the Jacobi coordinates $\mathbf{x}$, $\mathbf{y}$, and the center-of-mass coordinate $\mathbf{R}$. Under this transformation, the Hamiltonian~(\ref{H0}) takes the form
\begin{equation}
H_{0}=H_{r}+H_{R}+H_{xyR},  
\label{XYR}
\end{equation}%
where
\begin{equation}
H_{r}=-\frac{1}{2\mu }\left( \nabla _{\mathbf{x}}^{2}+\nabla _{%
\mathbf{y}}^{2}\right) +\frac{e^{2}B^{2}}{8}\frac{\mu }{%
m_{1}+m_{2}+m_{3}}\left[\left( \frac{m_{1}}{m_{2}^{2}}+\frac{m_{2}}{m_{1}^{2}}+%
\frac{m_{3}}{m_{1}^{2}}+\frac{m_{3}}{m_{2}^{2}}-\frac{m_{3}}{m_{1}m_{2}}%
\right) x^{2}+\left( \frac{m_{1}}{m_{3}^{2}}+\frac{m_{2}}{m_{3}^{2}}+\frac{%
m_{3}}{m_{1}m_{2}}\right) y^{2}\right],  
\label{Ho}
\end{equation}

\begin{equation}
H_{R}=-\frac{1}{2\mu }\nabla _{\mathbf{R}}^{2}+\frac{e^{2}B^{2}}{8}\frac{\mu }{m_{1}+m_{2}+m_{3}}\left( \frac{1}{m_{1}}+\frac{1}{m_{2}}+
\frac{1}{m_{3}}\right) R^{2}  
\label{HR}
\end{equation}
The Hamiltonian (\ref{XYR}) has three Laplace operators with respect to the
Jacobi coordinates, three terms proportional to the square of the Jacobi
coordinates $x^{2}$, $y^{2}$, and $R^{2}$, and the term $H_{xyR}$.  
The Hamiltonian (\ref{Ho}) describes the internal motion of three non-interacting charged particles with different masses in a magnetic field, while \( H_{R} \) corresponds to the motion of their $c.m.$ The term \( H_{xyR} \) captures the entanglement between the internal and collective degrees of freedom, 
and, therefore, prevents the separation of the $c.m.$ and relative motion in the general case of unequal masses. This term \( H_{xyR} \) consists of three contributions involving mixed Jacobi coordinates \( xy \), \( xR \), and \( yR \)—each multiplied by a factor determined by distinct algebraic combinations of the masses \( m_1 \), \( m_2 \), and \( m_3 \). The explicit expression for \( H_{xyR} \) is given in the Appendix, Eq.~(\ref{HXYR}). A careful examination of the mass factors in front of \( xy \), \( xR \), and \( yR \) in Eq.~(\ref{HXYR}) reveals that \( H_{xyR} = 0 \) in the equal-mass case. 
Furthermore, analyzing the expression for of the mass factors in front of \( x^2 \), \( y^2 \), and \( R^2 \) in (\ref{Ho}) and (\ref{HR})
shows that they become equal, namely \( \frac{1}{\sqrt{3} m} \), when all three particles have equal mass \( m_1 = m_2 = m_3 = m \). 
In this case, these Hamiltonians describe different degrees of freedom, and the corresponding wavefunctions depend on different variables: \(H_{r}\) acts on the internal coordinates \((x,y)\), whereas \(H_{\mathrm{R}}\) acts on the center-of-mass coordinate \(\mathbf{R}\). It is noteworthy that, for a system of three noninteracting equal-mass charged particles, the Hamiltonian $H_r$ separates into a sum of harmonic-oscillator Hamiltonians whose general solutions are expressed in terms of confluent hypergeometric functions. The energy quantization condition forces the confluent hypergeometric series to terminate, yielding Hermite polynomials. Consequently, the eigenfunctions of \(H_r\) factorize as $\phi(\mathbf{x})\,\chi(\mathbf{y}),
$ where \(\phi(\mathbf{x})\) and \(\chi(\mathbf{y})\) depend independently on the Jacobi coordinates \(\mathbf{x}\) and \(\mathbf{y}\), respectively.
\begin{figure}[t]
\centering
\includegraphics[width=6.5cm]{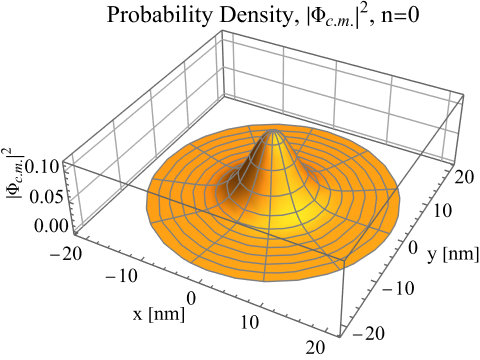}
\includegraphics[width=6.5cm]{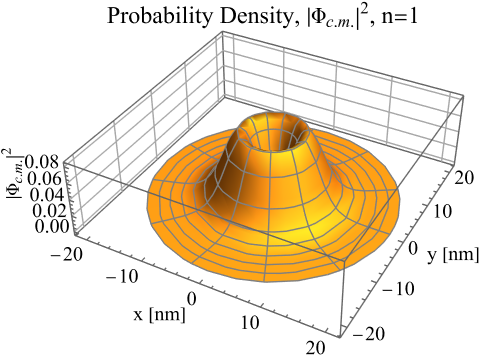}
\includegraphics[width=6.5cm]{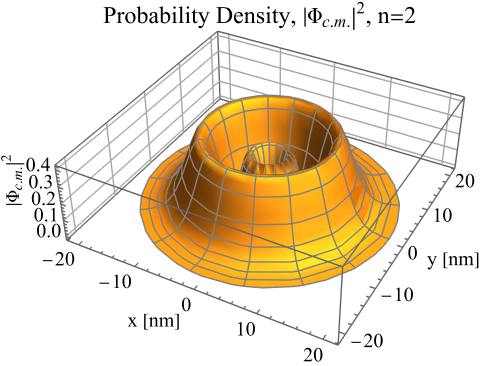}
\includegraphics[width=6.5cm]{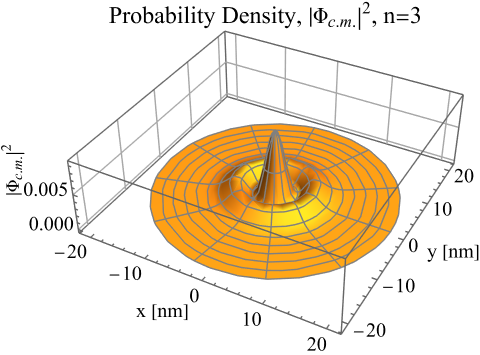}
\caption{(Color online) Three-dimensional plots of $|\Phi_{c.m.}|^2$ versus $x$ and $y$ of the the Center-of-Mass of three the same mass fermions in magnetic field $B=50$ T and electric field $E_{\perp} =0.5$ V/\AA in different states. 
}
\label{fig:CMWF}
\end{figure}

\begin{figure}[h!]
\centering
\includegraphics[width=17.0cm]{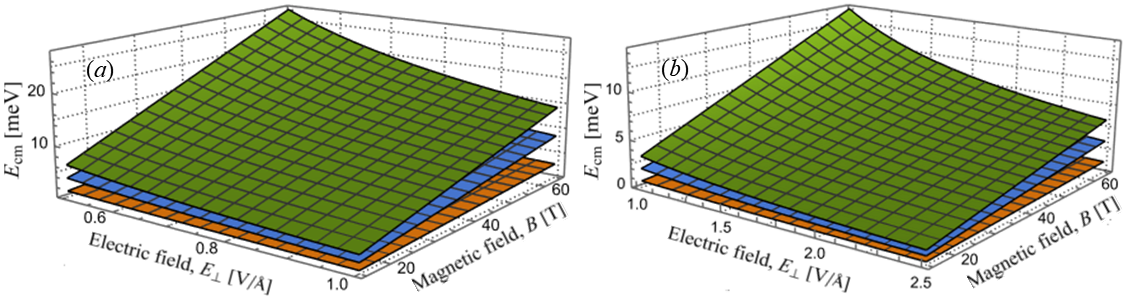}
\caption{(Color online) Center-of-mass Landau energy surfaces of the $X^{\mp}$ trions in silicene monolayer for the lowest three Landau levels as functions of the external electric and magnetic fields. Panels ($a$) and ($b$) show the results for the electric fields interval $0.5\leq E_\perp \leq 1.0$~V/\AA  and $1.0 \leq E_\perp \leq 2.5$~V/\AA, respectively. 
}
\label{LandauSurface}
\end{figure}

Therefore, in the case of three noninteracting particles trapped in a magnetic field with equal masses, the Hamiltonian (\ref{XYR}) 
decouples into a part 
describing the internal motion and a part 
corresponding to the $c.m.$ motion. The corresponding Schr\"{o}dinger equation for the center-of-mass reads:

\begin{equation}
-\frac{\sqrt{3}}{2m}\nabla _{\mathbf{R}}^{2}\Phi (\mathbf{R)}+%
\frac{e^{2}B^{2}}{8\sqrt{3}m}R^{2}\Phi (\mathbf{R)}=E_{\text{c.m.}}\Phi
(\mathbf{R)}\ .  \label{EHO}
\end{equation}
This equation is formally identical with the Schr\"{o}dinger equation in 1D space for a linear oscillator with the effective mass $m/\sqrt{3}$, oscillating with frequency $\varpi =\frac{1}{2}\frac{eB}{m}$ (the oscillator length, 
which determines the spatial width of the eigenfunctions, differs by a factor of \(3^{1/4}\) due to the different effective mass) 
and can be solved exactly with the energy spectrum
\begin{equation}
E_{\text{c.m.}}=\varpi \left(\mathfrak{N}+\frac{1}{2}\right), 
\label{Energy_cm}
\end{equation}
where $\mathfrak{N} = 0,1,2,3,...$ 
Equation (\ref{EHO}) with eigenvalues (\ref{Energy_cm}) yields the quantized $c.m.$ energy levels for a system of three equal-mass fermions confined within a magnetic trap.

The cyclotron motion of the $c.m.$ generates a time-dependent motional electric field in the center-of-mass frame, which couples to the internal degrees of freedom and generally prevents an exact separation of the center-of-mass and internal motions in charged systems~\cite{Schmelcher1991}. In contrast, we show that intervalley trions in Xene monolayers with equal electron and hole effective masses represent an exceptional case: for three equal-mass particles, the Hamiltonian admits an exact separation of the center-of-mass and internal motions.

We calculate the probability distributions of the $c.m.$ motion and quantized Landau energy surfaces of the $c.m.$ as a function of the external magnetic and electric fields for intervalley trions in Xene monolayers. Here and below, calculations are performed for $X^\mp$ trions formed by electrons and holes belonging to the valley-spin $A$ branches. The results for \(X^\mp\) trions formed from the $B$ branches are very close to those obtained for the $A$ branch trions because, for electric fields above $E_\perp \approx 0.6,\mathrm{V}/\text{\AA}$, the differences between the quasiparticle masses in the considered materials (Si, Sn, and Ge) are small.
 
Figure~\ref{fig:CMWF} illustrates the probability density of the center-of-mass wave function,
$|\Phi_{\mathrm{c.m.}}|^2$, for the four lowest quantized states of three equal-mass particles confined by a magnetic field. The ground state ($n=0$) exhibits the expected Gaussian-like distribution with a maximum at the origin, whereas the excited states develop concentric nodal rings characteristic of a two-dimensional harmonic oscillator. As the quantum number increases, the probability density extends over a larger spatial region, reflecting the increasing spatial extent of the center-of-mass motion. These results demonstrate that the center-of-mass of the three-particle complex undergoes Landau-like quantization induced solely by the magnetic field.

Figure~\ref{LandauSurface} presents the quantized Landau surfaces of the center-of-mass energy for intervalley trions in Xene monolayers as functions of the perpendicular electric and magnetic fields. Each surface corresponds to a different Landau quantum number and represents a quantized center-of-mass state of the trion. To highlight the effect of the electric field, the results are shown separately for two electric-field intervals: ($a$) $0.5 \leq E_\perp \leq 1.0$~V/\AA\ and ($b$) $1.0 \leq E_\perp \leq 2.5$~V/\AA. In both regimes, increasing the magnetic field strengthens the magnetic confinement and shifts all Landau surfaces to higher energies, while the electric field modifies the effective carrier masses and, consequently, the energies and spacing of the Landau surfaces. The electric-field dependence is significantly stronger in the low-field region, where the effective masses of the electron and hole are very small and 
whereas the Landau surfaces become less sensitive to the electric field at larger $E_\perp$. These results demonstrate that the center-of-mass motion of a trion is quantized into electrically tunable Landau states, whose energies can be controlled by the simultaneous application of external electric and magnetic fields independently of the trion internal binding energy. 

The stronger separation of the Landau surfaces observed in Fig.~\ref{LandauSurface}(a) is not a direct consequence of the electric field itself, but rather of the strong electric-field dependence of the effective carrier masses in Xenes. Since the Landau-level spacing is proportional to the cyclotron frequency, $\Delta E=\omega=\frac{ eB}{2m^{*}(E_\perp)}$, the electric field modifies the Landau spectrum through its effect on the effective mass. In the low-field region, close to the critical electric field, the effective masses vary rapidly with $E_\perp$, leading to a pronounced change in the cyclotron frequency and, consequently, a larger separation between adjacent Landau surfaces. At higher electric fields, however, the effective masses increase more gradually with $E_\perp$, causing the cyclotron frequency to vary more slowly. As a result, the Landau surfaces become nearly parallel and their spacing exhibits a much weaker dependence on the electric field, as shown in Fig.~\ref{LandauSurface}(b).

\section{Internal motion of three particles in trions in Xene monolayers}

\label{InternalMotion}

The exact separation obtained in the preceding section relies on a unique combination of symmetry conditions: (i) equal electron and hole effective masses, which result in identical cyclotron frequencies; (ii) a uniform perpendicular magnetic field; (iii) quadratic magnetic confinement in the symmetric gauge; and (vi) an interaction potential that depends only on the relative coordinates of the particles, as is the case for the Rytova–Keldysh potential. Under these conditions, the three-body Hamiltonian factorizes exactly into independent $c.m.$ and internal Hamiltonians, enabling separate descriptions of the collective Landau quantization and the internal trion dynamics.

Having established the separation of the $c.m.$ and internal motions for equal-mass three noninteracting particles in an external magnetic field for intervalley trions, we now turn to the internal dynamics of the three-particle system. We assume equal electron and hole effective masses and start from the Hamiltonian for three noninteracting particles in an external magnetic field, Eq.~(\ref{Ho}). The interparticle interactions are incorporated by adding the potential term
$
\sum_{i=1}^{3}V_i\!\left(\left|\mathbf{x}_i\right|\right),$
which yields the Schr\"odinger equation governing the internal motion of the interacting three-particle system:
\begin{equation}
\left[
-\frac{\sqrt{3}}{2m}
\left(
\nabla_{\mathbf{x}}^{2}
+
\nabla_{\mathbf{y}}^{2}
\right)
+
\frac{e^{2}B^{2}}{8\sqrt{3}m}
\left(
x^{2}+y^{2}
\right)
+
\sum_{i=1}^{3}
V_i\!\left(\left|\mathbf{x}_i\right|\right)
\right]
\Psi(\mathbf{x}_i,\mathbf{y}_i)
=
E\Psi(\mathbf{x}_i,\mathbf{y}_i).
\label{Relative3}
\end{equation}

To obtain a solution of the Schr\"{o}dinger equation (\ref{Relative3}) for the negatively and positively charged trions, we use the method of
hyperspherical harmonics (HH) \cite{Avery}. The main idea of this method is
the expansion of the wave function of the trion in terms of \ HH that are
the eigenfunctions of the angular part of the Laplace operator in the
four-dimensional (4D) space. In Eq. (\ref{Relative3}), $V_{i}(\left\vert 
\mathbf{x}_{i}\right\vert )$ is the interaction potential between two
particles 
at the relative distance $x_{1}$, $x_{2}$, and $x_{3}$, respectively, where $%
x_{i}$ is the modulus of the Jacobi vector $\mathbf{x}_{{i}}$ (\ref{Jacobi3}%
), and (\ref{Relative3}) is written for any of set $i=1,2,3$ of the Jacobi
coordinates (\ref{Jacobi3}). The orthogonal transformation between three
equivalent sets of the Jacobi coordinates 
simplifies calculations of matrix elements involving $V_{i}(\left\vert 
\mathbf{x}_{i}\right\vert )$ potentials.

The hyperspherical harmonics method has been widely used to describe three- and four-particle systems in two-dimensional materials and quantum dots~\cite{Ruan2000JPCM,Xie2000EPJB,Xie2001PSSB,FilikhinKez,Kezerashvili2008FBS,KezerashviliTsiklauri2013,KezerashviliTsiklauri2017FBS}. More recently, this method was successfully applied to the study of trions in Xene monolayers~\cite{Kezerashvili2024PRB}. Since the formalism has been presented in detail in Ref.~\cite{KezPRB109trion}, only a brief outline is given here. We introduce in the 4D space the hyperradius $\rho =\sqrt{%
x_{i}^{2}+y_{i}^{2}}$ and a set of three angles $\Omega _{i}\equiv (\alpha
_{i},\varphi _{x_{i}},\varphi _{y_{i}}),$ where $\varphi _{x_{i}}$ and $%
\varphi _{y_{i}}$ are the polar angles for the Jacobi vectors $\mathbf{x}%
_{i} $ and $\mathbf{y}_{j},$ respectively, and $\alpha _{i}$ is an angle
defined as $x_{i}=\rho \cos \alpha _{i},$ $y_{i}=\rho \sin \alpha _{i}$.
Next, we rewrite the Schr\"{o}dinger equation (\ref{Relative3}) for the
trion using hyperspherical coordinates in the 4D configuration space \cite%
{KezPRB109trion}. This transformation allows to reduce the solution of the
problem for the three particles in the 2D configuration space to the motion
of one particle in the 4D configuration space. Then we introduce the
hyperspherical harmonics $\Phi _{K\lambda }(\Omega )$ in the 4D
configuration space, which are the eigenfunctions of the angular part of the
generalized Laplace operator $\widehat{K}^{2}(\Omega _{i})$ in the 4D
configuration space $\widehat{K}^{2}(\Omega _{i})\Phi _{K\lambda }(\Omega
)=K(K+2)\Phi _{K\lambda }(\Omega )$ \cite{Avery}, where $K$ is a grand
angular momentum. Here we are using the short-hand notation $\lambda \equiv $
$\{l_{x},l_{y},L,M\},$ where $L$ is the total orbital angular momentum of
the trion, $M$ is its projection, and the grand angular momentum $%
K=2n+l_{x}+l_{y}$, $l_{x}$, where $l_{y}$ are angular momentum corresponding
to $\mathbf{x}$ and $\mathbf{y}$ Jacobi coordinates, respectively, and $n$ $%
\geqslant 0$ is an integer number.

The functions $\Phi _{K\lambda }(\Omega )$ present a complete set of
orthonormal basis, and one can expand the wave function of the trion $\Psi
(\rho ,\Omega _{i})$ in terms of the HH $\Phi _{K\lambda }(\Omega )$ as 
\begin{equation}
\Psi (\rho ,\Omega _{i})=\rho ^{-3/2}\sum_{_{K\lambda }}u_{K\lambda }(\rho
)\Phi _{K\lambda }(\Omega _{i}).\   \label{ExpanTrion}
\end{equation}%
In Eq. (\ref{ExpanTrion}) $u_{K\lambda }(\rho )$ are the hyperradial
functions and by substituting (\ref{ExpanTrion}) into the Schr\"{o}dinger
equation written in the hyperspherical coordinates \cite{ KezPRB109trion,Kezerashvili2024PRB}, one
can separate the radial and angular variables and get a set of coupled
differential equations for the hyperradial functions $u_{K\lambda }(\rho )$:

\begin{equation}
\left[ \frac{d^{2}}{d\rho ^{2}}-\frac{(K+1)^{2}-1/4}{\rho ^{2}}-\frac{%
e^{2}B^{2}}{8}\frac{1}{3m}\rho ^{2}+
\kappa ^{2}\right] u_{K\lambda }(\rho )=\frac{2m}{\sqrt{3}}%
\sum_{_{K^{^{\prime }}\lambda ^{^{\prime }}}}\mathcal{W}_{K\lambda
K^{^{\prime }}\lambda ^{^{\prime }}}(\rho )u_{K^{^{\prime }}\lambda
^{^{\prime }}}(\rho ).  \label{TrionGeneral}
\end{equation}%
In Eq. (\ref{TrionGeneral}) $\kappa ^{2}=2mE_{b}/\sqrt{3},$ where $%
E_{b}$ is trion BE. The coupling effective potential energy $\mathcal{W}%
_{K\lambda K^{^{\prime }}\lambda ^{^{\prime }}}(\rho )$ is 
\begin{equation}
\mathcal{W}_{K\lambda K^{^{\prime }}\lambda ^{^{\prime }}}(\rho )=\int \Phi
_{K\lambda }^{\ast }(\Omega _{i})\sum_{i<j}^{3}V_{ij}(\left\vert \mathbf{r}%
_{i}-\mathbf{r}_{j}\right\vert )\Phi _{K^{^{\prime }}\lambda ^{^{\prime
}}}(\Omega _{i})d\Omega _{i}.  \label{W3general}
\end{equation}

The coupling effective interaction (\ref{W3general}) is defined via the RK
potential~(\ref{Keldysh}). Substituting (\ref{Keldysh}) into Eq. (\ref%
{W3general}), one obtains the matrix elements of the effective potential
energies. 
The method of calculations of the effective potential energies is given in 
\cite{KezPRB109trion}. Calculations of matrix elements $\mathcal{W}_{K\lambda
K^{^{\prime }}\lambda ^{^{\prime }}}(\rho )$ of the two-body $%
V_{ij}(\left\vert \mathbf{r}_{i}-\mathbf{r}_{j}\right\vert )$ interactions
in the hyperspherical harmonics expansion method for a three-body system is
greatly simplified by using the HH basis states appropriate for the
partition corresponding to the interacting pair. Using the matrix elements $%
\mathcal{W}_{K\lambda K^{^{\prime }}\lambda ^{^{\prime }}}(\rho )$ in Eq. (%
\ref{TrionGeneral}), one can solve the system of coupled differential
equations numerically.

\begin{figure}[h!]
\centering
\includegraphics[width=8.5cm]{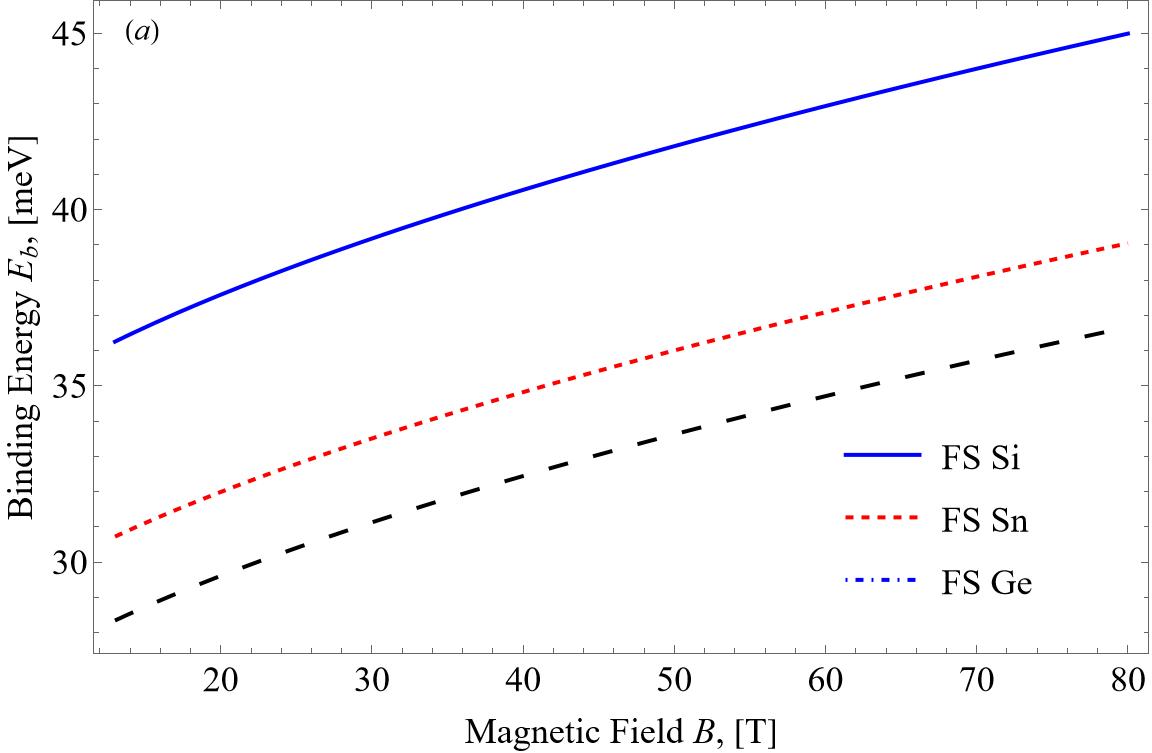}
\includegraphics[width=8.5cm]{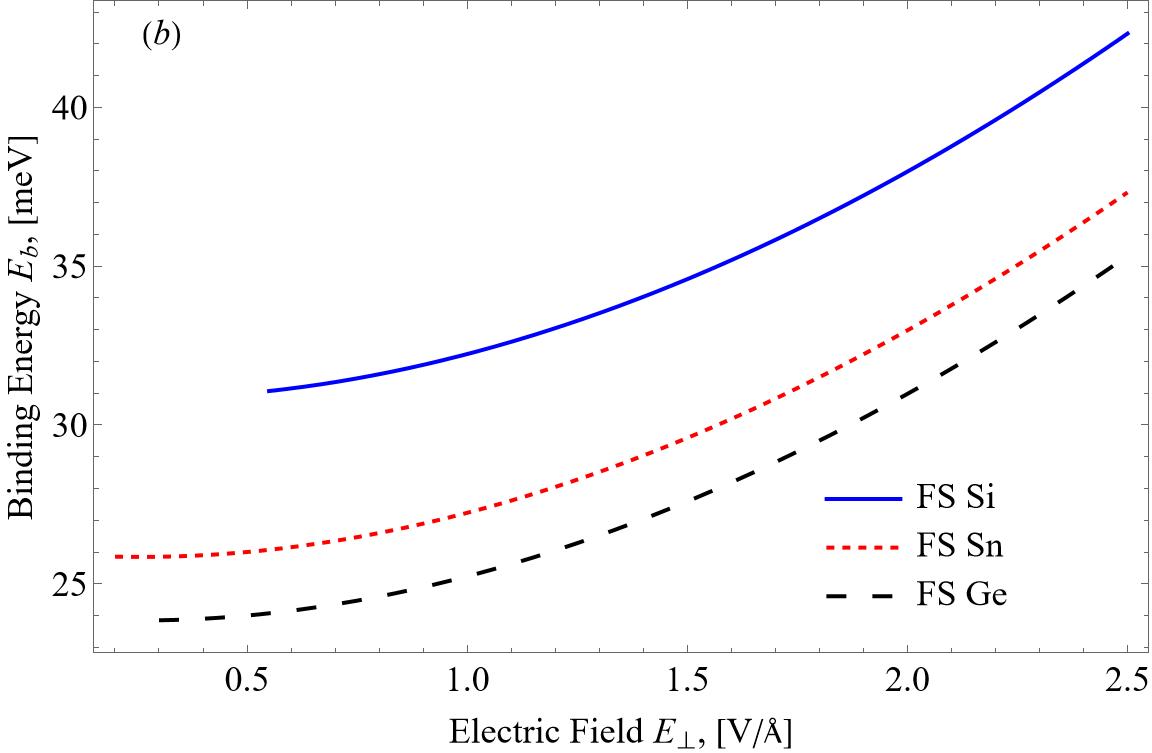}
\caption{(Color online) 
Dependence of the binding energy of intervalley $X^\mp$ magnetotrions in free-standing silicene, germanene, and stanene. ($a$) Binding energy as a function of the magnetic field $B$ for a fixed perpendicular electric field $E_\perp=2$~V/\AA. ($b$) Binding energy as a function of the perpendicular electric field $E_\perp$ for a fixed magnetic field $B=40$~T. The results demonstrate that both magnetic confinement and the electric-field-induced modification of the effective carrier masses enhance the trion binding energy. 
}
\label{BindingEnergyEM}
\end{figure}

\begin{figure}[h!]
\centering
\includegraphics[width=9.5cm]{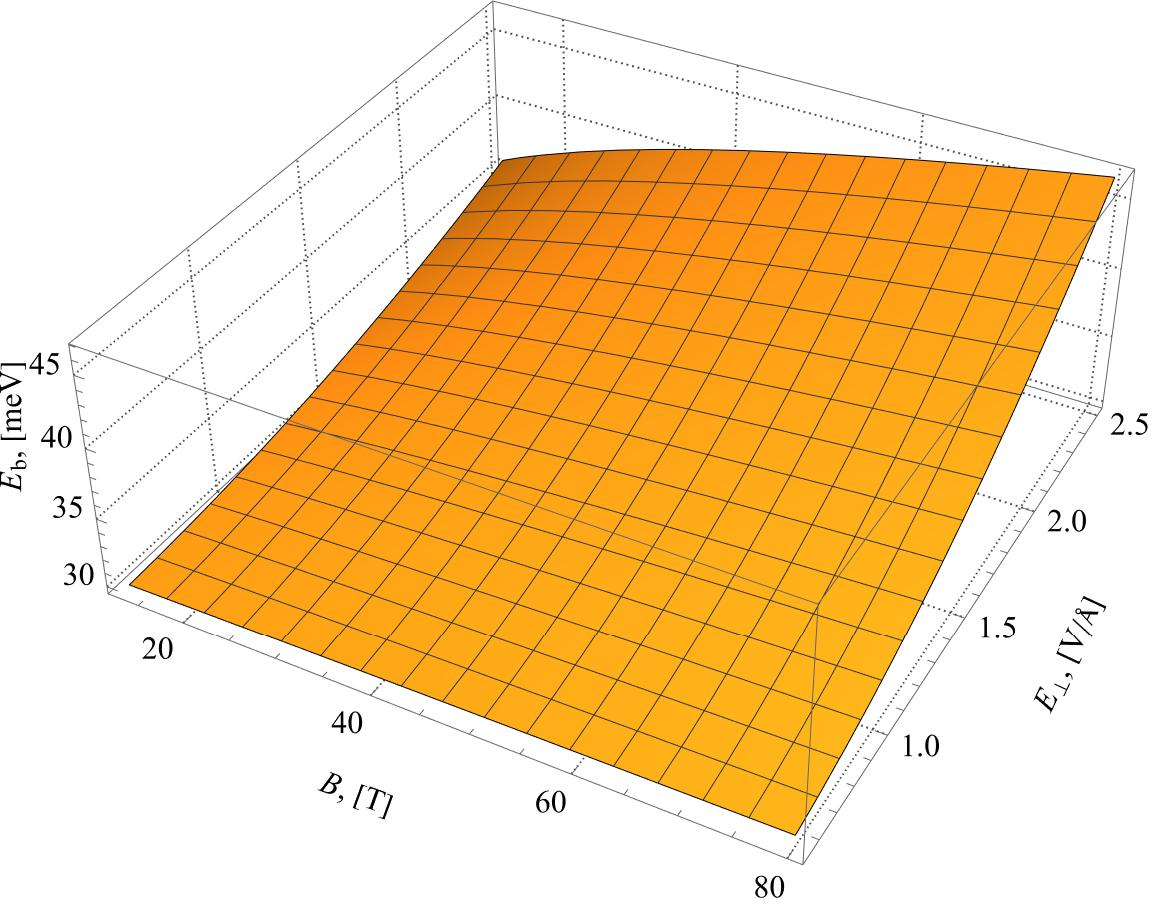}
\caption{(Color online) Binding-energy surface of the intervalley $X^{\mp}$ magnetotrion in a free-standing silicene monolayer as a function of the perpendicular magnetic field $B$ and electric field $E_\perp$. The surface illustrates the combined effect of magnetic confinement and the electric-field-dependent effective masses on the internal binding of the trion.}
\label{fig:EBSurface}
\end{figure}

We report the dependence of the intervalley magnetotrion binding energies in free-standing Xene monolayers (silicene, germanene, and stanene) on the external magnetic and perpendicular electric fields. The calculations were performed using the material parameters listed in Table~\ref{Table1}. Following  Ezawa~\cite{Ezawa3}, we restrict our analysis to the regime in which the applied perpendicular electric field exceeds the critical value, $E_{\perp}>E_c$, up to $E_\perp =2.5$ V/\AA. The corresponding critical electric fields for free-standing Xene monolayers are given in Ref.~\cite{Brunetti2019_BK}. The numerical solutions of Eq.~(\ref{W3general}) for intervalley trions are presented in Figs.~\ref{BindingEnergyEM} and~\ref{fig:EBSurface}.

\begin{table}[b]
\caption{Parameters of free-standing Xene monolayers used to calculate the effective masses and trion binding energies. Here, $2\Delta_{\rm so}$ is the total band gap between the conduction and valence bands, $d_0$ is the buckling parameter, $v_F$ is the Fermi velocity, $l$ is the monolayer thickness, and $\chi_{\rm 2D}=l\epsilon/4\pi$ is the two-dimensional polarizability.}
\label{Table1}
\begin{ruledtabular}
\begin{tabular}{lccc}
 & Si FS & Ge FS & Sn FS \\
\hline
$2\Delta_{\rm so}$ (meV)
& 1.9 (1.55)~\cite{Matthes2013a,Liu2011}
& 33 (23.9)~\cite{Matthes2013a,Liu2011}
& 101 (73.5)~\cite{Matthes2013a,Liu2011PR}
\\
$d_0$ (\AA)
& 0.46~\cite{Ni2012}
& 0.676~\cite{Ni2012}
& 0.85~\cite{Matthes2013a}
\\
$v_F$ ($\times10^{5}$ m/s)
& 6.5 (5.3)~\cite{Matthes2013a,Matthes2013PRB}
& 6.2 (5.2)~\cite{Matthes2013a,Matthes2013PRB}
& 5.5 (4.8)~\cite{Matthes2013a,Liu2011PR}
\\
$l$ (nm)
& 0.40~\cite{Tao2015}
& 0.45
& 0.50
\\
$\chi_{\rm 2D}$ (\AA)
& 3.788
& 5.730
& 9.549
\\
\end{tabular}
\end{ruledtabular}
\end{table}

Figure~\ref{BindingEnergyEM} shows the dependence of the binding energy of intervalley $X^\mp$ magnetotrions in free-standing silicene, germanene, and stanene on the external magnetic and electric fields. Panel~(a) demonstrates that the trion binding energy increases monotonically with the magnetic field for all three Xene monolayers. This behavior reflects the enhanced magnetic confinement of the three charge carriers, which compresses the internal wave function and increases the Coulomb interaction, resulting in more strongly bound trion states. The increase is most pronounced in silicene, followed by stanene and germanene, indicating that the magnetic-field response is strongly influenced by the material-dependent effective masses.

Figure \ref{BindingEnergyEM}(b) illustrates the dependence of the trion binding energy on the perpendicular electric field for a fixed magnetic field of $B=40$~T. The binding energy increases nonlinearly with $E_\perp$ in all three materials because the electric field modifies the band gap and, consequently, the effective masses of the charge carriers. Larger effective masses reduce the kinetic energy of the three-particle system and strengthen the Coulomb correlations, leading to an enhanced trion binding. Throughout the investigated field range, silicene exhibits the largest binding energy, whereas germanene possesses the smallest, demonstrating that the internal properties of intervalley trions can be efficiently controlled by the simultaneous application of external magnetic and electric fields.

We now extend the analysis from the one-dimensional cuts presented in Fig.~\ref{BindingEnergyEM}, corresponding to fixed values of $E_\perp$ or $B$, to the complete two-dimensional parameter space. Figure~\ref{fig:EBSurface} presents the binding-energy surface of the intervalley $X^{\mp}$ magnetotrion as a function of the perpendicular electric and magnetic fields. In contrast to Fig.~5, where the magnetic- and electric-field dependences are shown separately, this figure illustrates their combined influence on the trion internal binding. The binding energy increases monotonically with both external fields, reflecting the cooperative action of magnetic confinement and the electric-field-induced increase of the effective carrier masses. The magnetic field enhances the localization of the three-particle wave function, thereby strengthening the Coulomb interaction, whereas the electric field increases the effective masses through the field-dependent band gap, reducing the kinetic energy of the carriers. As a result, the binding-energy surface exhibits a smooth monotonic increase throughout the investigated parameter range, demonstrating that the internal stability of intervalley trions can be continuously tuned by the simultaneous application of perpendicular electric and magnetic fields.

The HH calculations demonstrate that the internal properties of intervalley trions can be efficiently controlled by external electric and magnetic fields. The numerical results show that both the magnetic and perpendicular electric fields enhance the trion binding energy through different physical mechanisms. The magnetic field increases the localization of the three-particle wave function via magnetic confinement, whereas the electric field increases the effective carrier masses through the field-dependent band gap, thereby reducing the kinetic energy and strengthening the Coulomb correlations. Consequently, the combined action of these external fields enables continuous tuning of the internal stability of intervalley trions, making Xene monolayers a promising platform for electrically and magnetically controlled charged excitonic complexes.

To the best of our knowledge, neither experimental measurements nor independent theoretical calculations of intervalley magnetotrions in buckled Xene monolayers are currently available for direct comparison with the present results. Consequently, quantitative benchmarking against alternative many-body approaches, such as Diffusion Quantum Monte Carlo or the Stochastic Variational Method, is not yet possible. Nevertheless, the hyperspherical harmonics method employed here has demonstrated accuracy and good convergence for two-dimensional trionic complexes and has been successfully validated for both TMDC monolayers \cite{FilikhinKez,KezFew2017,Kezerashvili2019,Kezerashvili2026,KezPRB109trion} and Xene monolayers in the absence of a magnetic field \cite{Kezerashvili2024PRB}. Therefore, the present calculations provide reliable predictions for intervalley magnetotrions in buckled Xene monolayers and establish a benchmark for future experimental and theoretical investigations.

Finally, we also address the influence of the dielectric environment and substrate effects on the present predictions. Although the calculations are performed for free-standing Xene monolayers to elucidate the intrinsic physical mechanisms, realistic experimental implementations will likely employ hexagonal boron nitride (hBN) encapsulation to stabilize these chemically reactive buckled materials. In such structures, the effective dielectric constant increases from the vacuum limit to approximately 3.4--4.5, whereas the intrinsic material parameters, including the buckling constant ($d_0$) and Fermi velocity ($v_F$), remain essentially unchanged because of the weak van der Waals coupling between the Xene and hBN layers. The increased dielectric screening modifies both the long-range Coulomb interaction and the screening length entering the Rytova--Keldysh potential, leading to a reduction of the absolute trion binding energies. However, because the dielectric environment acts uniformly throughout the monolayer, it neither lifts the valley or spin degeneracies nor alters the carrier mass-degeneracy condition ($m_e = m_h$) that constitutes the central symmetry requirement for the exact separation of the $c.m.$ and internal motions. Consequently, while quantitative binding energies depend on the dielectric environment, the exact factorization of the Hamiltonian and the predicted electrically tunable Landau quantization of the $c.m.$ motion remain unaffected. Experimentally, these predictions should be accessible through magneto-optical spectroscopy of high-quality hBN-encapsulated Xene monolayers under strong perpendicular magnetic fields and electrostatic gating, using techniques analogous to those successfully applied to excitonic complexes in other two-dimensional semiconductors.


\section{Concluding remarks}

\label{Conclusion}
In this work, we have established that intervalley magnetotrions in Xenes with equal electron and hole effective masses constitute an exceptional charged three-body system in which the center-of-mass and internal motions separate exactly, enabling independent control of their collective and internal dynamics by external electric and magnetic fields.

We have developed a theoretical description of intervalley magnetotrions in buckled Xene monolayers subjected to perpendicular electric and magnetic fields. Within the effective-mass approximation, the three-particle Schr\"odinger equation was formulated using the Rytova--Keldysh interaction potential and analyzed in the high-magnetic-field regime. Particular attention was devoted to the fundamental problem of separating the $c.m.$ and internal motions in a magnetic field. 
Building on the general theory of magnetic translations, it is demonstrated that while an exact separation is impossible for generic charged systems, intervalley trions in Xenes with equal electron and hole effective masses constitute an exceptional case. Specifically, under quadratic magnetic confinement in the symmetric gauge, the total Hamiltonian factorizes exactly into independent $c.m.$ and internal parts.

The separated $c.m.$ Hamiltonian was shown to be equivalent to a one-dimensional harmonic oscillator in a magnetic trap, leading to quantized Landau states of the trion $c.m.$. We calculated the corresponding probability distributions and obtained electrically tunable Landau energy surfaces, demonstrating that the collective motion of the trion can be controlled independently of its internal dynamics. The Landau surfaces exhibit a strong dependence on the magnetic field and a pronounced electric-field tunability originating from the electric-field dependence of the effective carrier masses in Xene monolayers.

The internal motion of intervalley trions was investigated by solving the three-body Schr\"odinger equation within the framework of the hyperspherical harmonics method. The influence of a perpendicular magnetic field on the binding energy of the three-body complex is governed by the balance between kinetic, Coulomb, and
magnetic confinement effects. Numerical calculations performed for free-standing silicene, germanene, and stanene show that the trion binding energy increases monotonically with both the magnetic and electric fields. The magnetic field enhances the localization of the three-particle wave function through magnetic confinement, whereas the perpendicular electric field increases the effective carrier masses by modifying the band gap, thereby strengthening the Coulomb correlations. Throughout the investigated parameter range, silicene exhibits the largest binding energies, followed by stanene and germanene. 

The present work establishes a unified theoretical framework describing both the collective center-of-mass motion and the internal dynamics of intervalley magnetotrions in Xene monolayers. The predicted electrically tunable Landau states together with the controllable trion binding energies provide new opportunities for manipulating charged excitonic complexes in two-dimensional buckled materials and may be relevant for future magneto-optical, valleytronic, and quantum-device applications.

\appendix

\section{Appendix}

\subsection{Particles have different masses and the same electric charges}
\begin{eqnarray}
H_{xyR} &=&\frac{e^{2}B^{2}}{8}\left\{ 2\frac{\mu ^{2}}{%
(m_{1}+m_{2}+m_{3})}\left( \frac{1}{m_{1}^{2}}-\frac{1}{m_{2}^{2}}-\frac{%
m_{3}}{m_{1}m_{2}^{2}}+\frac{m_{3}}{m_{2}m_{1}^{2}}\right) xy+\right.  
\notag \\
&&2\frac{\mu }{(m_{1}+m_{2}+m_{3})^{3/2}}\sqrt{\frac{m_{1}m_{2}}{m_{1}+m_{2}}%
}\left( \frac{1}{m_{1}}-\frac{1}{m_{2}}+\frac{m_{2}}{m_{1}^{2}}-\frac{m_{1}}{%
m_{2}^{2}}+\frac{m_{3}}{m_{1}^{2}}-\frac{m_{3}}{m_{2}^{2}}\right) xR+  \notag
\\
&&\left. 2\frac{\mu }{(m_{1}+m_{2}+m_{3})^{3/2}}\sqrt{\frac{%
(m_{1}+m_{2})m_{3}}{m_{1}+m_{2}+m_{3}}}\left( \frac{1}{m_{1}}+\frac{1}{m_{2}}%
-\frac{1}{m_{3}}-\frac{m_{1}}{m_{3}^{2}}-\frac{m_{2}}{m_{3}^{2}}+\frac{m_{3}%
}{m_{1}m_{2}}\right) yR\right\}   
\label{HXYR}
\end{eqnarray}

The transformation from the Cartesian coordinates $\mathbf{r}_{1}$, $\mathbf{%
r}_{2}$, and $\mathbf{r}_{3}$ to the Jacobi coordinates $\mathbf{x}$, $%
\mathbf{y}$, and $\mathbf{R}$ for the sum of $\frac{r_{i}^{2}}{m_{i}},$ $%
i=1,2,3$ gives

\begin{eqnarray}
\frac{r_{1}^{2}}{m_{1}}+\frac{r_{2}^{2}}{m_{2}}+\frac{r_{3}^{2}}{m_{3}} &=&%
\frac{\mu }{m_{1}+m_{2}+m_{3}}\left( \frac{m_{1}}{m_{2}^{2}}+\frac{m_{2}}{%
m_{1}^{2}}+\frac{m_{3}}{m_{1}^{2}}+\frac{m_{3}}{m_{2}^{2}}-\frac{m_{3}}{%
m_{1}m_{2}}\right) x^{2}+  \notag \\
&&\frac{\mu }{m_{1}+m_{2}+m_{3}}\left( \frac{m_{1}}{m_{3}^{2}}+\frac{m_{2}}{%
m_{3}^{2}}+\frac{m_{3}}{m_{1}m_{2}}\right) y^{2}+  \notag \\
&&\frac{\mu }{m_{1}+m_{2}+m_{3}}\left( \frac{1}{m_{1}}+\frac{1}{m_{2}}+\frac{%
1}{m_{3}}\right) R^{2}+  \notag \\
&&2\frac{\mu ^{2}}{(m_{1}+m_{2}+m_{3})}\left( \frac{1}{m_{1}^{2}}-\frac{1}{%
m_{2}^{2}}-\frac{m_{3}}{m_{1}m_{2}^{2}}+\frac{m_{3}}{m_{2}m_{1}^{2}}\right)
xy+  \notag \\
&&2\frac{\mu }{(m_{1}+m_{2}+m_{3})^{3/2}}\sqrt{\frac{m_{1}m_{2}}{m_{1}+m_{2}}%
}\left( \frac{1}{m_{1}}-\frac{1}{m_{2}}+\frac{m_{2}}{m_{1}^{2}}-\frac{m_{1}}{%
m_{2}^{2}}+\frac{m_{3}}{m_{1}^{2}}-\frac{m_{3}}{m_{2}^{2}}\right) xR+  \notag
\\
&&2\frac{\mu }{(m_{1}+m_{2}+m_{3})^{3/2}}\sqrt{\frac{(m_{1}+m_{2})m_{3}}{%
m_{1}+m_{2}+m_{3}}}\left( \frac{1}{m_{1}}+\frac{1}{m_{2}}-\frac{1}{m_{3}}-%
\frac{m_{1}}{m_{3}^{2}}-\frac{m_{2}}{m_{3}^{2}}+\frac{m_{3}}{m_{1}m_{2}}%
\right) yR  
\label{allm}
\end{eqnarray}

The presence of the terms $xy$, $xR$\ and $yR$ in Eq. (\ref{allm}) does not allow
for decoupling the $c.m.$ and internal motions.

When we have two particles with equal masses $m_{1}=m_{2}\equiv m$ and $%
m_{3}$ is the mass of the third particle the structure of the quadratic form (\ref{allm}) simplifies considerably:
\begin{eqnarray}
\frac{r_{1}^{2}}{m}+\frac{r_{2}^{2}}{m}+\frac{r_{3}^{2}}{m_{3}}&=&\mu_{(12)}\left[\frac{1}{m^{2}}\,x^{2}
+
\frac{1}{2m+m_3}
\left(
\frac{2m}{m_3^{2}}
+
\frac{m_3}{m^{2}}
\right)y^{2}
+
\frac{1}{2m+m_3}
\left(
\frac{2}{m}
+
\frac{1}{m_3}
\right)R^{2}\right]
\notag \\
&&+\frac{2\sqrt{2}\,\mu\left(M^{2}-m^{2}\right)}
{M^{3/2}m^{3/2}(2m+M)}yR
\label{12}
\end{eqnarray}
with 
\begin{equation}
\mu_{(12)}
=
\sqrt{\frac{m^{2}m_3}{2m+m_3}}.
\end{equation}

Thus, for the physically relevant trion case with two identical particles,
$m_1=m_2$, 
the mixed \(xy\) and \(xR\) coupling terms vanish identically, leaving only the \(yR\) coupling.

The case $m_1=m_3\equiv m$ and the third mass particle has the mass $m_2$, Eq. (\ref{allm}) becomes:
\begin{eqnarray}
\frac{r_{1}^{2}}{m}
+\frac{r_{2}^{2}}{m_{2}}
+\frac{r_{3}^{2}}{m}
&=&
\mu_{(13)}
\left[
\frac{1}{2m+m_2}
\left(
\frac{2m}{m_2^{2}}
+\frac{m_2}{m^{2}}
+\frac{1}{m}
-\frac{1}{m_2}
\right)x^{2}
\right.
\nonumber\\[2mm]
&&+
\frac{1}{2m+m_2}
\left(
\frac{1}{m}
+\frac{m_2}{m^{2}}
+\frac{1}{m_2}
\right)y^{2}
\nonumber\\[2mm]
&&\left.
+
\frac{1}{2m+m_2}
\left(
\frac{2}{m}
+\frac{1}{m_2}
\right)R^{2}
\right]
\nonumber\\[2mm]
&&+
\frac{2\mu_{(13)}^{2}(m_2-m)}
{m_2^{2}m^{2}}\,xy
\nonumber\\[2mm]
&&+
\frac{2\mu_{(13)}(m_2-m)\sqrt{m_2+m}}
{m_2^{3/2}m^{3/2}\sqrt{2m+m_2}}\,xR
\nonumber\\[2mm]
&&-
\frac{2\mu_{(13)}(m_2-m)\sqrt{m_2+m}}
{m_2m^{3/2}(2m+m_2)}\,yR.
\label{13}
\end{eqnarray}
with
\begin{equation}
\mu_{(13)}
=
\sqrt{\frac{m^{2}m_2}{2m+m_2}}.
\end{equation}
In the case $m_2=m_3\equiv m$ and the mass of the third particle is $m_1$, from Eq. (\ref{allm}), we obtain:
\begin{equation}
\begin{aligned}
\frac{r_{1}^{2}}{m_{1}}
+\frac{r_{2}^{2}}{m}
+\frac{r_{3}^{2}}{m}
={}&
\mu_{(23)}
\Bigg[
\frac{1}{2m+m_{1}}
\left(
\frac{m_{1}+m}{m^{2}}
+
\frac{2m-m_{1}}{m_{1}^{2}}
\right)x^{2}
\\
&\quad+
\frac{1}{2m+m_{1}}
\left(
\frac{m_{1}+m}{m^{2}}
+
\frac{1}{m_{1}}
\right)y^{2}
\\
&\quad+
\frac{m+2m_{1}}
{(2m+m_{1})m m_{1}}\,R^{2}
\Bigg]
\\[1mm]
&-
\frac{2\mu_{(23)}^{2}(m_{1}-m)}
{m_{1}^{2}m^{2}}\,xy
\\
&-
\frac{2\mu_{(23)}(m_{1}-m)\sqrt{m_{1}+m}}
{m_{1}^{3/2}m^{3/2}\sqrt{2m+m_{1}}}\,xR
\\
&-
\frac{2\mu_{(23)}(m_{1}-m)\sqrt{m_{1}+m}}
{m_{1}m^{3/2}(2m+m_{1})}\,yR .
\end{aligned}
\label{23}
\end{equation}
with
\begin{equation}
\mu_{(23)}
=
\sqrt{\frac{m^{2}m_1}{2m+m_1}}.
\end{equation}
The three special cases considered above correspond to different choices of the particle carrying the unequal mass. Consequently, the mass-scaled Jacobi coordinates are defined differently in each case, leading to different algebraic forms of the coefficients multiplying the quadratic terms $x^{2}$, $y^{2}$, $R^{2}$, $xy$, $xR$, and $yR$. These differences are purely kinematic and originate from the coordinate transformation rather than from any change in the underlying physics. The three expressions (\ref{12}), (\ref{13}), and (\ref{23}) are related by a permutation of the particle labels and therefore describe equivalent three-body dynamics expressed in different Jacobi coordinate systems. In the equal-mass limit, $m_1=m_2=m_3$, the permutation symmetry is fully restored, all three expressions reduce to the same form,

\begin{equation}
\frac{r_{1}^{2}}{m}+\frac{r_{2}^{2}}{m}+\frac{r_{3}^{2}}{m}=
\frac{1}{\sqrt3\,m}
\left(
x^2+y^2+R^2
\right).
\end{equation}

and the distinction between different Jacobi coordinate sets disappears.


\end{document}